\documentclass[aps,pre,reprint,superscriptaddress,longbibliography,groupedaddress,nofootinbib]{revtex4-2}

\usepackage{graphicx} % Required for inserting images
\usepackage{amsmath}
\usepackage{bm}
\usepackage{amssymb}
\usepackage{dsfont}

\usepackage{verbatim}
\usepackage[T1]{fontenc}
\usepackage[english]{babel}
\usepackage[utf8]{inputenc}

\begin{document}

\title{Ultimate Kinetic Uncertainty Relation 
\\and Optimal Performance of Stochastic Clocks
}

\author{Katarzyna Macieszczak}
\affiliation{Department of Physics, University of Warwick, Coventry CV4 7AL, United Kingdom}

\begin{abstract}
For Markov processes over discrete configurations, an asymptotic bound on the uncertainty of stochastic fluxes is derived in terms of the harmonic mean of decay rates with respect to the stationary distribution. This bound is necessarily tighter than the bound in terms of the arithmetic mean, i.e., the activity,  known as the kinetic uncertainty relation.  What is more, it can always be saturated. In turn, an exact limit for the uncertainty of first-passage times as well as the optimal long-time performance of stochastic clocks are established.  The results generalise to semi-Markov processes, including quantum reset processes, where it can be determined when clock performance improves thanks to coherent driving.
\end{abstract}

\maketitle

\emph{Introduction}. Systems with dynamics described by Markov processes are generally found away from equilibrium, with the detailed balance conditions not satisfied at long times due to the presence of asymptotic currents~\cite{Seifert2012}. Exact results for thermodynamic properties of such systems are rare and include fluctuations theorems~\cite{Jarzynski1997,Hatano2001,Seifert2005}, while many, such as fluctuation-dissipation theorem~\cite{Kubo1966}, Onsager~\cite{Onsager1931} or Green-Kubo relations~\cite{Green1954,Kubo1957}, are valid typically in linear response regime near equilibrium (cf., e.g., Refs.~\cite{Prost2009,Baiesi2009,Andrieux2007,Seifert2010}). Even in the context of equilibrium dynamics, the second thermodynamic only bounds the entropy change to be non-negative, the fact actually implied by Jarzynski equality~\cite{Jarzynski1997}, with similar inequalities originating from any fluctuation theorem. 

Recently, a different type of bounds, so called thermodynamic~\cite{Barato2015,Gingrich2016,Dechant2018} and kinetic uncertainty relations~\cite{Garrahan2017} have been uncovered, for the asymptotic ratio of the variance and squared average of integrated stochastic currents and fluxes, respectively. Those relations have been since extended to finite times~\cite{Horowitz2017,Dechant2018a,Terlizzi2019,Ito2020,Liu2020}, first-passage times~\cite{Garrahan2017,Gingrich2017,Hiura2021}, as well as, generalised from classical to quantum dynamics~\cite{Carollo2019,Hasegawa2020,Carollo2021,Vu2022} and in certain settings  linked to fluctuation theorems~\cite{Hasegawa2019,Timpanaro2019}. They do not, however, generally saturate for any choice of currents or fluxes, as system observables need to be considered as well~\cite{Koyuk2020,Shiraishi2021}. For example, the thermodynamic uncertainty relation is can saturate asymptotically only in the linear response regime, which occurs for the entropy production current~\cite{Macieszczak2018,Shiraishi2021}. 

In this work, we show that the bound in the kinetic uncertainty relation (KUR)~\cite{Garrahan2017}, which is given by the activity, can be improved in the long-time limit so that it can actually be saturated, and thus the ultimate asymptotic KUR is established. The improved bound is given by the inverse of the harmonic rather than the arithmetic mean of the decay rates with respect to the stationary distribution, and coincides with the average observed lifetime of system configurations rather than their activity. As a consequence, we find necessary and sufficient condition for the original KUR to be saturated asymptotically: the decay rates need to be uniform, in which case the KUR saturates for the activity flux. Furthermore, thanks to the asymptotic correspondence between generalised ensembles of trajectories~\cite{Budini2014}, the derived uncertainty relation also establishes the ultimate uncertainty relation for first passage times in the asymptotic limit~\cite{Garrahan2017}, i.e., statistics of times when the integrated flux crosses a given extensive value. 

Beyond the fundamental importance, the new bound has an operational interpretation in terms of performance of stochastic clocks~\cite{Milburn2020}.  While both of the uncertainty of stochastic fluxes (or currents~\cite{Barato2016}) as well as first-passage times can be considered as figure of merits for the precision of time estimation, noise correlations in continuous clock operation are correctly accounted for by Allan variance~\cite{Allan1966}. Nevertheless, due to correlations effectively no longer contributing at long times, all those notations coincide, and the universal limit on performance of stochastic clocks is found.

Finally, our results generalise from Markov to semi-Markov processes, with the bound in so far known KUR~\cite{Carollo2019,Vu2020} replaced by the ultimate bound generally distinct from the average lifetime. The semi-Markov processes describe stochastic dynamics of classical configurations with memory, but they also capture stochastic trajectories of so called quantum reset or renewal dynamics. For such open quantum systems, we demonstrate that  Hamiltonian driving can improve stochastic clock performance but optimal estimation in general requires counting subsequent transitions (cf.~Refs.~\cite{Cilluffo2022,Manikandan2023}).

\emph{Ultimate KUR}.  For an ergodic Markov process, a time-integrated stochastic flux is $K(t)= \int_{0}^t dt'\sum_{jk} n_{jk}(t') \,\alpha_{jk}$, where $n_{jk}(t)$ takes value $1$ if a transition from  $j$ to $k$ configurations occurs at time $t$ and $0$ otherwise, while $\alpha_{jk}$ are real numbers that define the choice of the flux. The flux uncertainty is given by the ratio of the variance $\Delta^2 K(t)$ to the square of its squared average $\langle K(t)\rangle$.  Our first main result is that the uncertainty can be asymptotically bounded as 
\begin{align}\label{eq:hKUR}
\lim_{t\rightarrow\infty} t\, \frac{\Delta^2 K(t)}{\langle K(t)\rangle^2 }\geq	\sum_j  \frac{p_j^\text{ss}}{\lambda_j}  
 \geq \frac{1}{\sum_j p_j^\text{ss}   \lambda_j} . 
\end{align}
Here, for a system configuration $j$, $p_j^\text{ss}$ denotes the stationary probability and $\lambda_j$ is the decay rate, which can be expressed as $\lambda_j\equiv\sum_{k} w_{jk}$ for  transition rates $w_{jk}$ from $j$ to $k$ [the stationary distribution then obeys $\sum_{j} p_{j}^\text{ss} w_{jk}= p_{k}^\text{ss} \lambda_k$].    

The first inequality is proven in~\cite{SM} by using the exact result for large deviations~\cite{denHollander2000} of Markov processes, the so called level-2.5 rate function~\cite{Maes2008,Bertini2015}. As the probability of no transition occurring when in a configuration $j$ decays exponentially at the rate  $\lambda_j$, $1/\lambda_j$ is the lifetime of  $j$, and the first bound can be recognised as the average lifetime in the stationary distribution.

The second inequality, which corresponds to the known KUR, is typically interpreted as the average rate of all transitions, i.e., the activity. In Eq.~\eqref{eq:hKUR}, it simply arises as a consequence of a weighted harmonic mean being greater than the corresponding weighted arithmetic mean, $1/(\sum_j  p_j^\text{ss}/\lambda_j ) \leq \sum_j p_j^\text{ss} \lambda_j$. From Eq.~\eqref{eq:hKUR}, it follows that the asymptotic saturation of KUR requires those means to coincide, which occurs if and only if the decay rates are uniform ($\lambda_j=\lambda$). This condition is thus necessary for the KUR to saturate at long times.

While asymptotically both $\langle K(t)\rangle$ and $\Delta^2 K(t)$ become linear in time $t$, the rate of the average is simply given by $\lim_{t\rightarrow\infty}  \langle K(t)\rangle/t =\sum_{jk} p_{j}^\text{ss} w_{jk}\alpha_{jk}$. The rate of variance   $\lim_{t\rightarrow\infty}  \Delta^2 K(t)/t =\sum_{jk} p_{k}^\text{ss} w_{jk}\alpha_{jk}^2 - 2 \sum_{jklm} p_{j}^\text{ss} w_{jk} \alpha_{jk}  s_{kl} w_{lm}   \alpha_{lm} $, also involves the dynamics resolvent with $\sum_{k}s_{jk} w_{kl}=\delta_{jl}-p_l^\text{ss}=\sum_{k}w_{jk} s_{kl}$, but, thanks to the uncertainty relations in Eq.~\eqref{eq:hKUR}, for any flux it is still bounded in terms of the stationary distribution and the transition rates.

\begin{figure*}[t!]
	\includegraphics[width=1.0\linewidth]{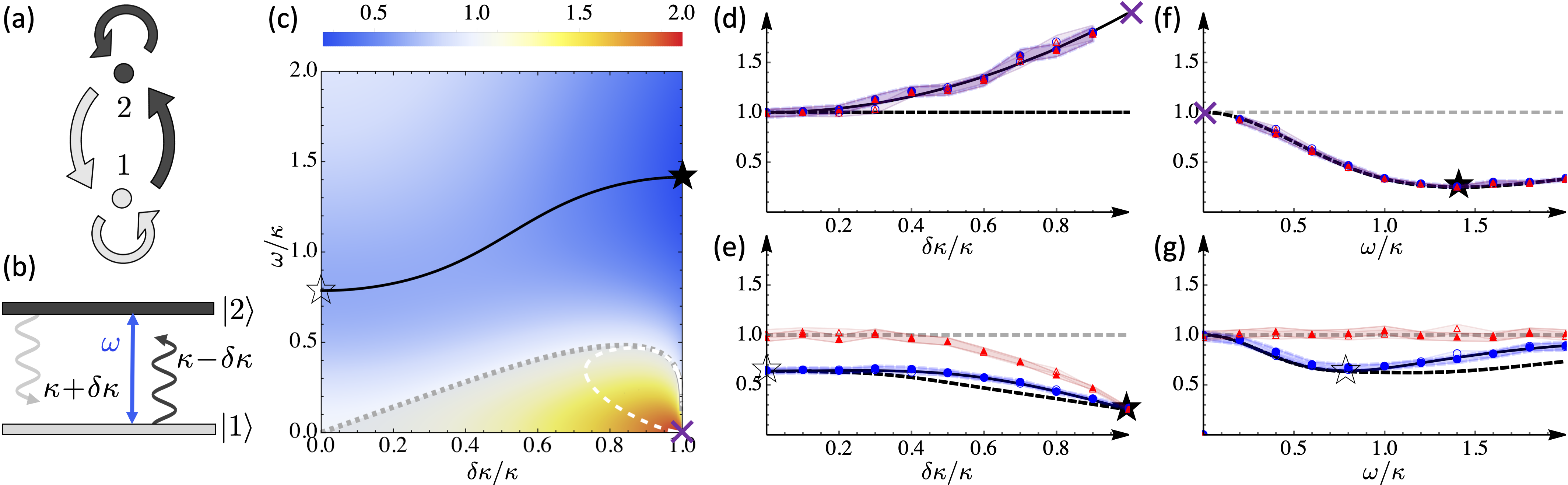}
	\caption{\label{fig}\textbf{From classical to quantum clocks}.
		\textbf{(a)} A semi-Markov process 
		for an open quantum system in \textbf{(b)}, which becomes Markov for $\omega=0$. 
		\textbf{(c)} The minimal cost-error product in the long-time limit, cf.~Eqs.~\eqref{eq:clock} and~\eqref{eq:clock_semi0}. The black solid line marks $\omega=\omega_\text{opt}(\delta \kappa)$ leading to the minimal value for given $\delta \kappa $. Above the dashed grey line, the classical limit is broken; outside the dashed white line this is allowed to happen by the known KUR, cf.~Eq.~\eqref{eq:hKUR_semi}; $\langle \tau\rangle_\text{ss}$ (not shown) is always lower than the ultimate KUR.  For $N(t)$ (red circles) and $K_\text{opt}^*(t)$ (blue triangles), the cost-error product (filled marker) is shown together with the uncertainty multiplied by the activity (empty markers) when  \textbf{(d)} $\omega/\kappa=0$, \textbf{(e)} $\omega=\omega_\text{opt}(\delta \kappa)$, \textbf{(f)}  $\delta\kappa/\kappa=0$, and \textbf{(g)}  $\delta\kappa/\kappa=1$. In (d)-(g), the classical limit is indicated as the grey dashed line, the semi-Markov bound in Eq.~\eqref{eq:clock_semi} as the black solid line, and the bound corresponding to the known KUR as the black dashed line.
		The data points were
		obtained from 5000 quantum trajectories each~\cite{Daley2014}, initialised from stationary distributions and 100 times longer than the configuration lifetimes ($T_{jj}\tau_{jj}+T_{jj\perp}\tau_{jj\perp}$), and the inverse of the gap  in the master operator ($1/\kappa$)~\cite{Lindblad1976,GKS1976}; 
		the error bars show twice the estimated standard deviation.
		In (c), (d), (f), the purple cross marks the limits $\omega/\kappa=0$ and $\delta \kappa/\kappa=1$.
		In (c), (e), (f), and (g), the empty and filled stars correspond to $\omega_\text{opt}(\delta \kappa)$ for $\delta \kappa/\kappa=0,1$, respectively. }
\end{figure*}

\emph{Optimal stochastic flux}. For the flux,
\begin{equation}\label{eq:K_opt}
K_\text{opt}(t)\equiv \int_{0}^t dt'\,\sum_{jk} n_{jk}(t') \,\frac{1}{\lambda_j},
\end{equation}
we show below that
\begin{equation}\label{eq:K_opt2}
    \lim_{t\rightarrow\infty} \frac{\langle K_\text{opt}(t)\rangle}{t} =1, \quad \lim_{t\rightarrow\infty} \frac{\Delta^2 K_\text{opt}(t)}{t}=\sum_j   \frac{p_{j}^{\text{ss}}}{\lambda_j}.
\end{equation}
It directly follows that for any real $\alpha\neq0$, the first inequality in Eq.~\eqref{eq:hKUR} saturates for $K_\alpha(t)=\alpha K_\text{opt}(t)$. Moreover, a choice of the stochastic flux with $\alpha_{jk}=\alpha/\lambda_{j}+(\beta_{j}-\beta_{k})$ with real $\beta_j$,  for which we will denote time-integral as $K_{\alpha,\vec{\beta}}(t)$, also leads to the equality in Eq.~\eqref{eq:hKUR} due to vanishing asymptotic total currents (that is, $\lim_{t\rightarrow\infty}\int_{0}^t dt'\,\sum_{k} [n_{jk}(t')-n_{kj}(t')]/t=0$ for all $j$~\cite{Bertini2015}). Importantly, this not only means that such fluxes are optimal, but that the uncertainty relation cannot be further improved upon, so Eq.~\eqref{eq:hKUR} is the ultimate KUR. Furthermore, when the decay rates are uniform ($\lambda_j=\lambda$), both inequalities in Eq.~\eqref{eq:hKUR} saturate, so this is not only necessary but sufficient for the asymptotic saturation of KUR. In that case, the total number of transitions, $N(t)\equiv\sum_{jk}\int_{0}^t dt'\sum_{jk} n_{jk}(t')$ with $\lim_{t\rightarrow\infty}\langle N(t)\rangle/t=\sum_j   p_{j}^{\text{ss}}{\lambda_j}$,  becomes an optimal choice of a time-integrated flux [as then $N(t)=K_{\lambda}(t)$, but it can be also optimal in other cases, cf.~Fig.~\ref{fig}].

Indeed, since $\alpha_{jk}=1/\lambda_j$ for $K_\text{opt}(t)$, $\lim_{t\rightarrow\infty}  \langle K_\text{opt}(t)\rangle/t=\sum_j(p_{j}^\text{ss} /\lambda_{j}) \sum_k w_{jk}=\sum_jp_{j}^\text{ss}=1$. Similarly, $\lim_{t\rightarrow\infty}  \Delta^2 K_\text{opt}(t)/t =\sum_{j} (p_{j}^\text{ss}/\lambda_j^2)\sum_k w_{jk} - 2 \sum_{j} (p_{j}^\text{ss}/\lambda_j)\sum_{kl}  w_{jk}   s_{kl} \sum_m w_{lm}   /\lambda_{l} =\sum_{j} p_{j}^\text{ss}/\lambda_j - 2 \sum_{j}  (p_{j}^\text{ss}/\lambda_j) \sum_{kl}  w_{kl}   s_{lm}=\sum_{j} p_{j}^\text{ss}/\lambda_j$ as $\sum_{kl}  w_{jk}   s_{kl}= \sum_{l}  (\delta_{jl}-p_l^\text{ss})=0$.

\emph{Ultimate first-passage time uncertainty}. For a chosen stochastic flux such that its asymptotic rate is non-zero, let $T(k)=t$ denote the shortest time $t$ when the time-integral $K(t)\geq k$ (we assume the rate to positive and thus $k>0$). Then, our second main result is that (cf.~Refs.~\cite{Garrahan2017,Hiura2021})
\begin{align}\label{eq:tKUR}
\lim_{k\rightarrow\infty}  \frac{\Delta^2 T(k)}{\langle T(k)\rangle }\geq	\sum_j  \frac{p_j^\text{ss}}{\lambda_j}  
 \geq \frac{1}{\sum_j p_j^\text{ss}   \lambda_j} . 
\end{align}
In particular, for the first-passage time $ T_\text{opt}(k)$ associated with $K_\text{opt}(t)$ in Eq.~\eqref{eq:K_opt}, we obtain 
\begin{equation}\label{eq:T_opt}
    \lim_{k\rightarrow\infty} \frac{\langle T_\text{opt}(k)\rangle}{k} = 1, \ \lim_{k\rightarrow\infty} \frac{\Delta^2 T_\text{opt}(k)}{k}=\sum_j   \frac{p_{j}^{\text{ss}}}{\lambda_j},
\end{equation}
and the uncertainty is saturated. This also occurs for first-passage time associated with $ K_{\alpha,\vec{\beta}}(t)$ and in~\cite{SM} is shown to follow from the direct relation between large deviations of fluxes and and their first-passage times~\cite{Budini2014,Gingrich2017}.
In contrast, for $T(n)=t$ associated with $N(t)\geq n$, in~\cite{SM} we show that $\lim_{n\rightarrow\infty} \langle T(n)\rangle/n=1/[\sum_j  p_{j}^{\text{ss}}\lambda_j]$ and  $\lim_{n\rightarrow\infty} \Delta^2 T(n)/n\geq [\sum_j  p_{j}^{\text{ss}}/\lambda_j]/[\sum_j   p_{j}^{\text{ss}}\lambda_j]$, so the inequality cannot saturate unless the decay rates are uniform. When they are, the choice of total number of transitions indeed becomes optimal with both the improved uncertainty and the known uncertainty saturating in Eq.~\eqref{eq:tKUR}.

\emph{Ultimate limit on Markov clock performance}. We now consider an application of our results to  stochastic clocks with Markov dynamics in the stationary operation limit, so that initial configurations can be assumed to be sampled from the stationary distribution. By measuring a stochastic flux, its time-integral $K(t)$ can be used to infer the elapsed time $t$. Indeed, $\langle K(t) \rangle= t  \langle K\rangle_\text{ss}$, where $\langle K\rangle_\text{ss}\equiv \sum_{jk} p_j^\text{ss} w_{jk}\alpha_{jk}$, assumed $\neq 0$, is the average flux for the system in the stationary distribution (which coincides with the asymptotic rate of $K(t)$ for any initial state). Therefore, $\overline{K}(t)\equiv K(t)/\langle K\rangle_\text{ss}$ is an unbiased estimator of $t$. The errors of time estimators are captured by the time Allen variance, which can be expressed as %$\Delta_A^2 K(2t,t)= \Delta^2 [K(t+2\tau)-2K(t+\tau)+K(t)]/(2\tau^2)=\Delta^2 [K(2\tau)-2K(\tau)]/(2\tau^2)$
$\Delta_A^2 \overline{K}[t,2t]\equiv \{\Delta^2 \overline{K}(t)+\Delta^2 [\overline{K}(2t)-\overline{K}(t)] -2\langle \overline{K}(t) ,\overline{K}(2t)-\overline{K}(t)\rangle\}/2=[\Delta^2 K(t)-\langle K(t) ,K(2t)-K(t)\rangle]/ \langle K\rangle_\text{ss}^2$, where  $\langle K(t),K(2t)-K(t) \rangle$ are the correlations between observations $K(t)$ and $K(2t)-K(t)$ in the subsequent time intervals of length $t$.

We next show that for the rate of the variance $\Delta^2 K(t)$ to be finite, as holds for ergodic dynamics, it is necessary that $\lim_{t\rightarrow\infty}\langle K(t),K(2t)-K(t) \rangle/t=0$. Indeed, we have $\Delta^2 K(2t)=\Delta^2 K(t)+\Delta^2 [K(2t)-K(t)]+2\langle K(t) ,K(2t)-K(t)\rangle= 2\Delta^2 K(t)+2 \langle K(t) ,K(2t)-K(t)\rangle$, so that $\lim_{t\rightarrow\infty} \Delta^2 K(t)/t=\lim_{t\rightarrow\infty}\Delta^2 K(2t)/(2t)=\lim_{t\rightarrow\infty} \Delta^2 K(t)/t+\lim_{t\rightarrow\infty} \langle K(t) ,K(2t)-K(t)\rangle/t$.
Importantly, it follows that
the rate of the time Allan variance rate asymptotically coincides with $\lim_{t\rightarrow\infty}  t \Delta^2 K(t)/\langle K(t)\rangle^2 $. Thus, from Eq.~\eqref{eq:hKUR}, it is bounded as
\begin{align}\label{eq:AKUR}
\lim_{t\rightarrow\infty} \frac{ \Delta_A^2  \overline{K}[t,2t]}{t}\geq	\sum_j  \frac{p_j^\text{ss}}{\lambda_j} 
\end{align}
and from Eq.~\eqref{eq:K_opt2},
\begin{equation}\label{eq:K_opt3}
   \lim_{t\rightarrow\infty}\frac{\Delta^2_A  K_\text{opt}[t,2t]}{t}=\sum_j   \frac{p_{j}^{\text{ss}}}{\lambda_j}.
\end{equation}
Furthermore, for any $\alpha\neq 0$, $\overline{K}_{\alpha}(t)=K_\text{opt}(t)$ and the limit for $K_{\alpha,\vec{\beta}}(t)$ is exactly same. As it can be saturated, the bound represents the ultimate limit on the stochastic clock performance, cf. Fig.~\ref{fig}.  Although the time Allan variance in the long-time limit increases linearly in time, the relative errors $\overline{K}[t,2t]/t^2$ 
 (the frequency Allan variance) decay inversely in time.

With the known best performance of a given stochastic clocks in Eq.~\eqref{eq:AKUR}, we can now consider the optimal choice of dynamics for time estimation, but this requires also quantifying the cost  running the clock. If the total activity is chosen to account for that, then the product of the cost and the relative errors, which we will refer to as the cost-error product, can be bounded in the long-time limit as
\begin{align}\label{eq:clock}
\lim_{t\rightarrow\infty} \langle N(t) \rangle \, \frac{\Delta_A^2  \overline{K}[t,2t]}{t^2}\geq	1. 
\end{align}
This again follows by the inequality between the arithmetic and harmonic means, which also implies that the cost-error product is minimised by the dynamics with the uniform decay rates.  This bound then is, for example, minimised by counting self-transitions for an only configuration $j$  (equivalent to the Poisson process at the rate $\lambda_j$), which among others describes radioactive decay exploited in radio carbon dating~\cite{Milburn2020}. Analogous results holds for the precision quantified in terms of the uncertainty of stochastic flux (and the same cost function) or the uncertainty of first-passage times (and the cost function being the average total number of transition for given $k$). While those bounds are equivalent to the known uncertainty relations, the optimal design is now clarified. Furthermore, below we show that the bound in Eq.~\eqref{eq:clock} can be further lowered for semi-Markov processes as the uncertainty relations are altered.

\emph{Upper bounds for finite times}. While the results so far are asymptotic, in~\cite{SM} we show that Eqs.~\eqref{eq:K_opt2} and~\eqref{eq:K_opt3} for any time $t$ provided that initial configurations are sampled from the stationary distribution. Then, the minimal uncertainty $ \Delta^2 K(t)/\langle K(t)\rangle^2$ and the minimal relative errors $
 \Delta_A^2 \overline{K}(t,2t)/t^2$ achieved with optimal choices of stochastic fluxes can be no worse than those for $K_\text{opt}(t)$ in Eq.~\eqref{eq:K_opt} and thus they are bounded from above by the average configuration lifetime divided by $t$, respectively.

\emph{Ultimate KUR for semi-Markov dynamics}. For a semi-Markov process, the exponential decay of the survival probability $s_j(\tau)$ of at least time $\tau$  spent in a configuration $j$ since the last transition, the so called age, is replaced by the decay with the age-dependent rate $\lambda_j(\tau)\equiv\sum_k w_{jk}(\tau)$. This is due to age-dependent transition rates $w_{jk}(\tau)\equiv
t_{jk}(\tau)/s_j(\tau)$, giving rise to $d s_j(\tau)/d\tau =-  \sum_k t_{jk}(\tau) $ with $s_j(0)=1$ [here, the dynamics is determined in terms $t_{jk}(\tau)$]. Assuming ergodicity, the asymptotic probability for a $j$ configuration at age $\tau$ is $p_j^\text{ss}(\tau)= c_j s_j(\tau)$, where $c_j>0$ fulfills $\sum_{j}c_j T_{jk} =c_k$ for $ T_{jk}\equiv \int_0^\infty d\tau \,t_{jk}(\tau)$ and it uniquely determined by the normalisation of the total probability. Then, the overall asymptotic probability of a configuration $j$ can be expressed as $c_j \int_0^\infty d\tau\,s_j(\tau)  = c_j \int_0^\infty d\tau\,\tau\,\sum_k t_{jk}(\tau) = c_j \sum_k T_{jk} \tau_{jk}$, where  $\tau_{jk}\equiv [\int_0^\infty d\tau\,\tau\, t_{jk}(\tau)]/T_{jk}$ is the average observed age of $j$ before a transition to $k$ (assumed $<\infty$ if $T_{jk}\neq 0$). Thanks to ergodicity,  $p_{jk}\equiv c_j  T_{jk} \tau_{jk}$ can also be identified as the average part of total time spent in configuration $j$ before transitioning to $k$. Furthermore, 
the rate of observed transitions from $j$ to $k$, $\int_0^\infty d\tau\,p_j^\text{ss}(\tau) w_{jk}(\tau)=c_j T_{jk}=p_{jk}/\tau_{jk}$, so that for an age-independent time-integrated stochastic flux $K(t)=\int_{0}^t dt'\,\sum_{jk} n_{jk}(t') \,\alpha_{jk}$, the asymptotic rate $\lim_{t\rightarrow\infty}\langle K(t)\rangle/t=\sum_{jk} p_{jk} \alpha_{jk}/\tau_{jk}\equiv\langle K \rangle_\text{ss}$. In particular, $\lim_{t\rightarrow\infty}\langle N(t)\rangle/t=\sum_{jk} p_{jk}/\tau_{jk}$.

In~\cite{SM}, by exploiting the level-2.5 rate function for semi-Markov processes~\cite{Sughiyama2018,Carollo2019}, we derive our third main result, which is  the following uncertainty relation
\begin{align}\label{eq:hKUR_semi}
\lim_{t\rightarrow\infty} t\, \frac{\Delta^2 K(t)}{\langle K(t)\rangle^2 }\geq	\sum_{jk} p_{jk} \frac{\sigma_{jk}^2}{\tau_{jk}} \geq \frac{1}{\sum_{jk} p_{jk} \frac{\tau_{jk}}{ \sigma_{jk}^2}},
\end{align}
where  $\sigma_{jk}\equiv [\int_0^\infty d\tau\,(\tau-\tau_{jk})^2\, t_{jk}(\tau)]/T_{jk}$ is the variance in age of $j$ before a transition to $k$ (assumed $<\infty$ if $T_{jk}\neq 0$). The second inequality, with the bound~\cite{Carollo2019} akin to the KUR but valid also for semi-Markov processes, follows from the first inequality again by considering the harmonic and arithmetic means with respect to $p_{jk}$.  
Importantly, for the stochastic flux
\begin{equation}\label{eq:K_opt_semi}
K^*_\text{opt}(t)\equiv \int_{0}^t dt'\,\sum_{jk} n_{jk}(t') \,\tau_{jk},
\end{equation}
in~\cite{SM} we show that
\begin{equation}\label{eq:K_opt_semi2}
    \lim_{t\rightarrow\infty} \frac{\langle K^*_\text{opt}(t)\rangle}{t} =1, \quad \lim_{t\rightarrow\infty} \frac{\Delta^2 K^*_\text{opt}(t)}{t}=\sum_{jk} p_{jk} \frac{\sigma_{jk}^2}{\tau_{jk}},
\end{equation}
so that the first inequality in Eq.~\eqref{eq:hKUR_semi} saturates. Thus the ultimate KUR for semi-Markov processes is established, see Fig.~\ref{fig}. The saturation also occurs for $K_{\alpha,\vec{\beta}}^*(t)$ with $\alpha_{jk}=\alpha \tau_{jk}+(\beta_{j}-\beta_{k})$ for real $\alpha\neq 0$ and $\beta_j$. We note, however, that while for uniform $\tau_{jk}$, an optimal flux can be chosen as total transition number $N(t)$, the bounds in Eq.~\eqref{eq:hKUR_semi} coincide if and only if the ratios $\sigma_{jk}^2/\tau_{jk}$ are uniform.

To connect with the results for the Markov case, the improved bound in Eq.~\eqref{eq:hKUR_semi} should be compared with the average lifetime, i.e., the average age in the asymptotic distribution, $\langle \tau\rangle_\text{ss}\equiv\sum_j \int_0^\infty d\tau\,\tau\, p_j^\text{ss}(\tau)=\sum_j c_j \int_0^\infty d\tau\,\tau^2\sum_k t_{jk}(\tau)/2=\sum_j c_j T_{jk} (\sigma_{jk}^2+\tau_{jk}^2)/2=\sum_j p_{jk} (\sigma_{jk}^2/\tau_{jk}+\tau_{jk})/2$. While those expressions coincide for any Markov process (where  $c_j=p^\text{ss}_j\lambda_{j}$, $T_{jk}=w_{jk}/\lambda_{j}$, $\tau_{jk}=1/\lambda_{j}$, and $\sigma^2_{jk}=1/\lambda_{j}^2$), no general relation exists for semi-Markov processes. Nevertheless, from  Eq.~\eqref{eq:K_opt_semi2}, the Markov-like uncertainty relation with $\langle \tau\rangle_\text{ss}$ as the bound on the uncertainty is valid but cannot be saturated if  $\sum_j p_{jk} \sigma_{jk}^2/\tau_{jk}> \sum_j p_{jk} \tau_{jk}$, while  that  relation can be invalided, e.g., by considering $K_{\alpha,\vec{\beta}}^*(t)$, when  $\sum_j p_{jk} \sigma_{jk}^2/\tau_{jk}< \sum_j p_{jk} \tau_{jk}$.

 We expect the bounds in Eq.~\eqref{eq:hKUR_semi} to hold also for the uncertainty of first-passage times as well as the improved bound to saturate when considering  $ T^*_\text{opt}(k)=t$ as the shortest time $t$ such that $K^*_\text{opt}(t)\geq k$, but we leave this as a hypothesis. In what remains, we focus on the operational interpretation of the ultimate KUR in Eq.~\eqref{eq:hKUR_semi}.

\emph{Ultimate limit on semi-Markov clock performance}. As in  Markov case, time-integrals of stochastic fluxes in semi-Markov dynamics can be used to estimate elapsed time $t$ by considering $\overline{K}(t)\equiv K(t)/\langle K\rangle_\text{ss}$  in the stationary operation regime. It then analogously follows from Eq.~\eqref{eq:hKUR_semi} that the rate of the time Allan variance is asymptotically bounded as
\begin{align}\label{eq:clock_semi}
\lim_{t\rightarrow\infty} \frac{\Delta_A^2 \overline{K}[t,2t]}{t }&\geq \sum_{jk} p_{jk}  \,\frac{\sigma_{jk}^2}{\tau_{jk}} ,
\end{align}
and the rate is minimal when considering $K_\text{op}^*(t)$ [or more generally $\overline{K}^*_{\alpha,\vec{\beta}}(t)$] in Eq.~\eqref{eq:K_opt_semi},
\begin{align}\label{eq:clock_semi0}
\lim_{t\rightarrow\infty} \frac{\Delta_A^2 K^*_\text{opt}[t,2t]}{t }&=\sum_{jk} p_{jk}  \,\frac{\sigma_{jk}^2}{\tau_{jk}}.
\end{align}
This leads to the cost-error product is bounded as
\begin{align}\label{eq:clock_semi1}
\lim_{t\rightarrow\infty} \langle N(t) \rangle \, \frac{\Delta_A^2 \overline{K}[t,2t]}{t^2 }&\geq \sum_{jk} p_{jk} \frac{1}{\tau_{jk}}\sum_{jk} p_{jk}  \,\frac{\sigma_{jk}^2}{\tau_{jk}} \\
\label{eq:clock_semi2}
&\geq \frac{\sum_{jk} p_{jk}  \frac{\sigma_{jk}^2}{\tau_{jk}} }{\sum_{jk} p_{jk} \,\tau_{jk}}. 
\end{align}
The first equality, Eq.~\eqref{eq:clock_semi1}, follows from Eq.~\eqref{eq:clock_semi}. If this bound is less than $1$, the cost-error product can be lower than the Markov limit in Eq.~\eqref{eq:clock}, and this is guaranteed  by Eq.~\eqref{eq:clock_semi0} to be achieved when considering $K_\text{opt}^*(t)$ [or $\overline{K}^*_{\alpha,\vec{\beta}}(t)$], see Fig.~\ref{fig}. %In contrast, by using the known KUR for semi-Markov, the obtained bound being lower than $1$ form only a necessary but not sufficient condition for improved clock performance. 
For example, counting self-transitions in the dynamics of an only configuration $j$ the Markov limit is broken if and only if $\sigma^2_{jj}< \tau^2_{jj}$, cf. Fig.~\ref{fig}. More generally, if $\sigma^2_{jk}< \tau^2_{jk}$ for all $j$ and $k$, the performance beyond the Markov can be achieved. Finally, the Markov limit can be actually broken to an arbitrary degree for any number of configurations in the dynamics. Indeed, 
$t_{jk}(\tau)$ that determines the semi-Markov dynamics can be defined in terms of $T_{jk}$ and the age cumulants, $\tau_{jk}$, $\sigma^2_{jk}$, .... If $T_{jk}$ and $\tau_{jk}$ are given for every $j$ and $k$, all $p_{jk}$ are determined, but $\sigma^2_{jk}$ can be varied independently and chosen arbitrarily small.

The second inequality, Eq.~\eqref{eq:clock_semi2}, again follows from the inequality between the harmonic and arithmetic means and it shows that the semi-Markov ultimate KUR must be lower than $\langle \tau\rangle_\text{ss}$, that is, the Markov ultimate KUR must be broken, for the clock performance to improve.

\emph{Example --- quantum reset process}. We finish by considering semi-Markov processes arising from quantum reset process and the limits on performance of corresponding clocks, cf.~Figs.~\ref{fig}(a) and~\ref{fig}(b). With the general correspondence discussed in~\cite{SM} (see also Ref.~\cite{Woods2022}), we focus on a $2$-level system. It resonantly driven at frequency $\omega$ and coupled to a thermal bath resulting in the decay from the state $j=1,2$ at rate $\kappa_j\equiv\kappa+(-1)^j\delta \kappa$, which resets the system to the opposite configuration $j^\perp\!\equiv j\! \!\mod 2\! +\!1$, see Fig.~\ref{fig}(b). In the classical limit, $\omega=0$, Markov dynamics with $w_{jj^\perp}=\kappa_j$ (and $w_{jj}=0$) are recovered. But for $\omega\neq 0$, transition rates are age-dependent, with $t_{j j}(\tau)=\kappa_{j^\perp}  |(e^{\tau \mathbf{G}})_{ jj^\perp}|^2$ and $t_{j j^\perp}(\tau)=\kappa_{j}  |(e^{\tau \mathbf{G}})_{j j}|^2$, where $(\mathbf{G})_{jj}\equiv\!-\kappa_j/2$ and $(\mathbf{G})_{ j j^\perp}\equiv\!-\!\mathsf{i}\omega/2 $, leading to $s_{j}(\tau)=|(e^{\tau \mathbf{G}})_{j j}|^2+|(e^{\tau \mathbf{G}})_{jj^\perp}|^2$. For the classical dynamics, only at infinite temperature  ($\delta \kappa/\kappa=0$) the decay rates are uniform so the standard KUR saturates. Otherwise. the lower the temperature (higher $\delta \kappa/\kappa$), the higher the cost-error product for the corresponding clock, but optimal time estimation can still be achieved by counting total number of transitions, see Figs.~\ref{fig}(c) and~\ref{fig}(d). At any temperature the clock performance can be improved by the presence of coherent drive, cf.~Fig.~\ref{fig}(c), and the classical limit of $1$ in Eq.~\eqref{eq:clock} can actually be broken by appropriately tuning the drive frequency, further allowing for the cost-error product to decrease with the temperature, see Fig.~\ref{fig}(e). Thus, the optimal dynamics is achieved at zero temperature ($\delta \kappa/\kappa=1$, where self-transitions of $1$ facilitated by resetting from $2$), with the limit $1/4$ attained at $\omega/\kappa=\sqrt{2}$, see Fig.~\ref{fig}(f) and cf.~Refs~\cite{Cilluffo2022,Manikandan2023}. Even at infinite temperature, the limit $0.64...$ can be achieved, but this requires sequential rather than cumulative counting (for the latter the total number of transitions is best but meets the classical limit), see Fig.~\ref{fig}(g). Interestingly, the cost-error product in not analytic in the limits of zero temperature and vanishing drive, and depends on their ordering, cf.~Figs. 1(d) and 1(f); this coincides with the dynamics becoming absorbing (non-ergodic).
Finally, the ultimate KUR is necessary to critically assess  whether the drive lowers the flux uncertainty or the relative errors at higher temperatures beyond the classical limit and thus the possibility of improved clock performance, cf.~Fig.~\ref{fig}(c).

\emph{Conclusions and outlook}.
This works introduces ultimate bounds on the asymptotic uncertainty relations for statistics of stochastic fluxes and their first-passage times in Markov processes. Those bounds are further linked to the long-time limit on the relative errors in time estimation, in turn allowing for considering optimal stochastic clocks. Those results generalise to  semi-Markov dynamics, among others, allowing for investigation of quantum stochastic clocks described by quantum reset processes. As these outcomes improve upon the known uncertainty relations, also identifying optimal fluxes, it comes as a natural question if this can be achieved also at finite times. It also remains to be seen whether the related bounds in general stochastic quantum dynamics arising from continuous monitoring of open quantum systems (cf.~Refs.~\cite{Hasegawa2020,Vu2022}) can be improved upon and what that would mean for quantum clock performance (cf.~Ref.~\cite{Meier2023}). 

\emph{Note}. Related results have appeared recently in Ref.~\cite{Prech2024}.

\bibliography{references.bib}

\end{document}